\documentclass{PoS}
\usepackage{url} 

\title{Direct $CP$ violation in charm at Belle}

\ShortTitle{Direct $CP$ violation in charm at Belle}

\author{\speaker{B. R. Ko}\thanks{on behalf of the Belle Collaboration.}\\
        Korea University, SEOUL, Republic of Korea\\
        E-mail: \email{brko@hep.korea.ac.kr}}

\abstract{Using the full data sample collected with the Belle detector
  at the KEKB asymmetric-energy $e^+e^-$ collider, we have searched
  for $CP$ violation of charmed mesons in $D^+\rightarrow K^0_S\pi^+$
  and $D^0\rightarrow h^+h^-$ decays, where $h$ denotes $K$ and
  $\pi$. We observe evidence for $CP$ violation in $D^+\rightarrow
  K^0_S\pi^+$ decay with 3.2 standard deviations away from zero,
  $(-0.363\pm0.094\pm0.067)\%$, while the asymmetry is consistent with
  the expected $CP$ violation due to the neutral kaon in the final
  state. No evidence for $CP$ violation in $D^0\rightarrow h^+h^-$ is
  observed with $A^{KK}_{CP}=(-0.32\pm0.21\pm0.09)\%$ and
  $A^{\pi\pi}_{CP}=(+0.55\pm0.36\pm0.09)\%$. The $CP$ asymmetry
  difference between $D^0\rightarrow K^+K^-$ and
  $D^0\rightarrow\pi^+\pi^-$ decays is also measured with $\Delta
  A^{hh}_{CP}=(-0.87\pm0.41\pm0.06)\%$, which is 2.1 standard
  deviations away from zero and supports recent LHCb and CDF
  measurements.}

\FullConference{36th International Conference on High Energy Physics,\\
		July 4-11, 2012\\
		Melbourne, Australia}

\begin{document}

Violation of the combined Charge-conjugation and Parity symmetries
($CP$) in the standard model (SM) is produced by a non-vanishing phase
in the Cabibbo-Kobayashi-Maskawa flavor-mixing matrix~\cite{CKM},
where the violation may be observed as a non-zero $CP$ asymmetry
defined as 
\begin{equation}
  A^{D\rightarrow f}_{CP}=\frac
  {\Gamma(D\rightarrow f)-\Gamma(\bar{D}\rightarrow\bar{f})}
  {\Gamma(D\rightarrow f)+\Gamma(\bar{D}\rightarrow\bar{f})}
  \label{EQ:ACP}
\end{equation}
where $\Gamma$ is the partial decay width, $D$ denotes a charmed
meson, and $f$ is a final state.

In this presentation, we report $CP$ asymmetries of charmed mesons in
the decays $D^+\rightarrow K^0_S\pi^+$, $D^0\rightarrow K^+K^-$,
$D^0\rightarrow\pi^+\pi^-$~\cite{CC}, and the $CP$ asymmetry
difference between $D^0\rightarrow K^+K^-$ and
$D^0\rightarrow\pi^+\pi^-$, which is an update of our previous
publications~\cite{ACPKSH, D0hhBelle} using the full data sample
collected with the Belle detector~\cite{BELLE} at the KEKB~\cite{KEKB}
asymmetric-energy $e^+e^-$ collider. The $D^+\rightarrow K^0_S\pi^+$
final state is a coherent sum of Cabibbo-favored and doubly
Cabibbo-suppressed decays where no SM $CP$ violation in charm decay is
expected, while $(-0.332\pm0.006)\%$~\cite{PDG2012} $CP$ violation due
to $K^0-\bar{K}^0$ mixing (denoted by $A^{\bar{K}^0}_{CP}$) is
expected with a neutral kaon in the final state.
The $D^0\rightarrow h^+h^-$ final states where $h$ denotes $K$ and
$\pi$ are singly Cabibbo-suppressed decays in which both direct
($a^{\rm dir}_{CP}$) and indirect $CP$ violations ($a^{\rm ind}_{CP}$)
are expected in the SM, while the $CP$ asymmetry difference between
the two decays, $\Delta A^{hh}_{CP}=A^{KK}_{CP}-A^{\pi\pi}_{CP}$,
reveals approximately direct $CP$ violation with the universality of
indirect $CP$ violation in charm decays~\cite{YAY}.
The data were recorded at the $\Upsilon(nS)$ resonances
($n=1,2,3,4,5$) or near the $\Upsilon(4S)$ resonance and the
integrated luminosity is $\sim$1 ab$^{-1}$.

We determine the quantity $A^{D\rightarrow f}_{CP}$ defined in
Eq.~(\ref{EQ:ACP}) by measuring the asymmetry in the signal yield
\begin{equation}
  A^{D\rightarrow f}_{\rm rec}~=~\frac{N_{\rm rec}^{D\rightarrow f}-N_{\rm rec}^{\bar{D}\rightarrow\bar{f}}}{N_{\rm
    rec}^{D\rightarrow f}+N_{\rm rec}^{\bar{D}\rightarrow\bar{f}}}  
  ~=~A^{D\rightarrow f}_{CP}~+~A_{FB}+~A^{f}_{\epsilon},
  \label{EQ:ARECON}
\end{equation}
where $N_{\rm rec}$ is the number of reconstructed decays. The
$A_{FB}$ is forward-backward asymmetry in $e^+e^-\rightarrow c\bar{c}$
process and the $A^{f}_{\epsilon}$ is final state particle detection
asymmetry where the latter depends on the final state particles while
the former does not. For a slow pion detection asymmetry which is
involved in $D^0\rightarrow h^+h^-$ reconstruction via $D^{*+}$, we
correct for the asymmetry using the method described in our previous
publication~\cite{D0hhBelle}. A fast pion detection asymmetry
which is involved in $D^+\rightarrow K^0_S\pi^+$ reconstruction is
corrected for using the method described in Ref.~\cite{KSPIPRL}. With
assumption the $A_{FB}$ is the same for all charmed mesons,
Refs.~\cite{D0hhBelle, KSPIPRL} use $CP$ violation free large
statistics of resonance data samples to correct for the $A^{f}_{\epsilon}$.
For the final state with a neutral kaon, we have to take into account
additional corrections which are asymmetry due to different
interactions between $K^0$ and $\bar{K}^0$ with detector~\cite{K0MAT}
and experiment dependent $A^{\bar{K}^0}_{CP}$ with $K^0_S$ decay time
dependency on it~\cite{GROSSMAN_NIR}. Once we correct for
$A^{f}_{\epsilon}$, then $A^{D\rightarrow f}_{CP}$ is obtained in bins
of the polar angle of charmed meson momentum at the center-of-mass
system (c.m.s.) using antisymmetry of $A_{FB}$ in the polar angle of
charmed meson momentum at the c.m.s. 

\begin{figure}[htbp]
\mbox{
  \includegraphics[width=0.35\textwidth, height=0.4\textwidth]{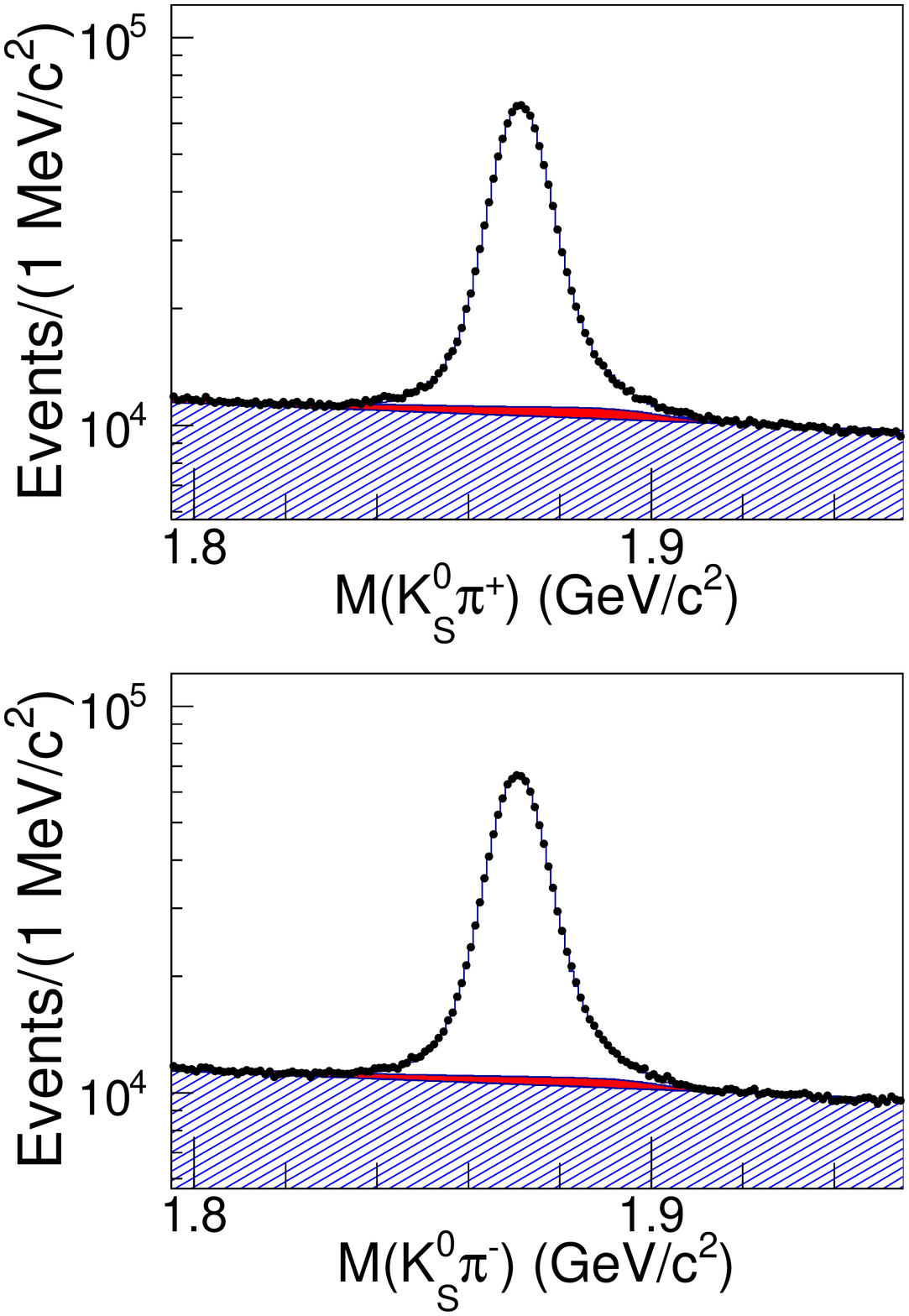}
}
\mbox{
  \includegraphics[width=0.65\textwidth, height=0.4\textwidth]{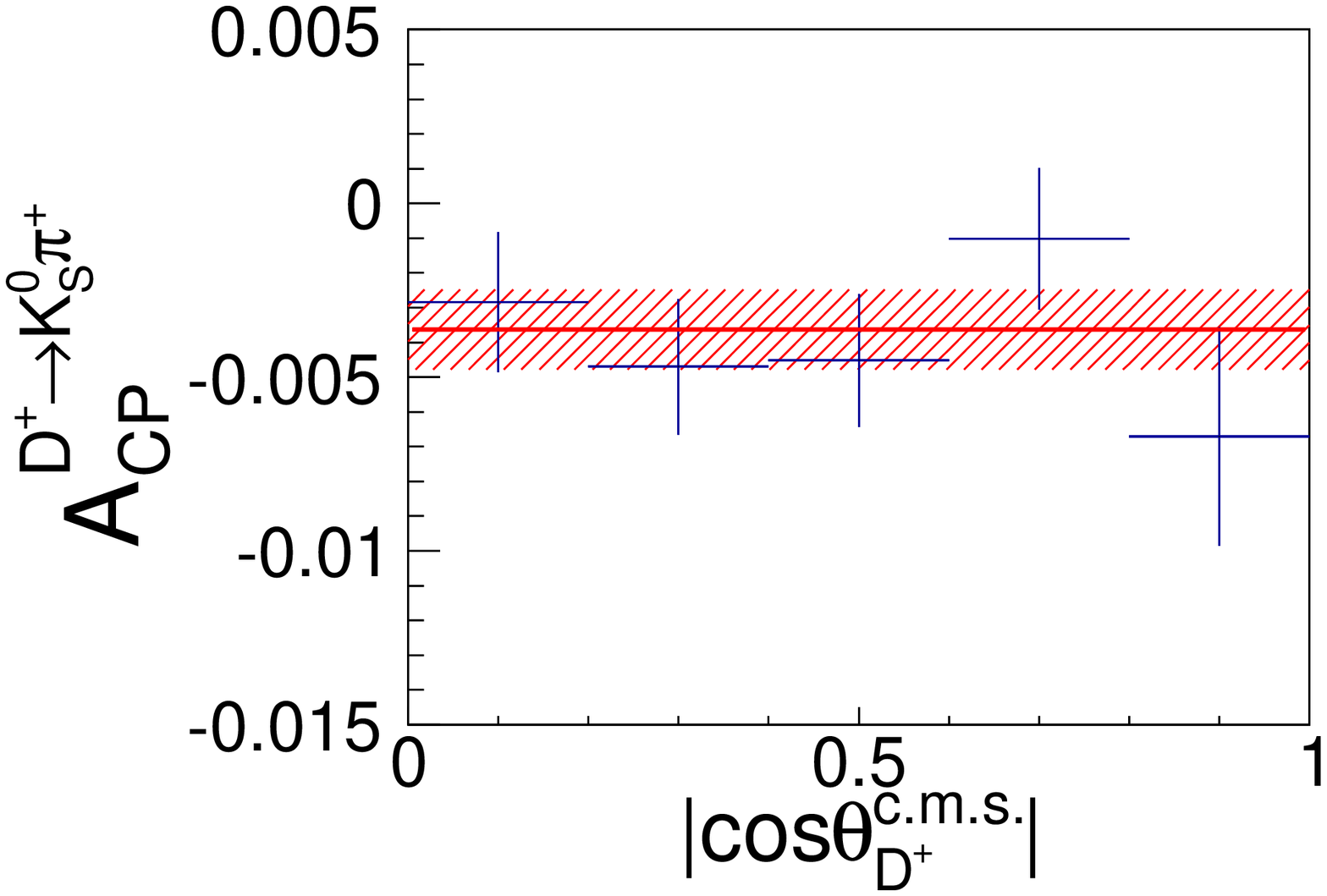}
}
\caption{$M(K^0_S\pi^+)$ (left top) and $M(K^0_S\pi^-)$ (left bottom)
  distributions where the shaded and hatched are $D^+_s\rightarrow
  K^0_S K^+$ due to particle misidentification and combinatorial
  backgrounds. Right plot is $A_{CP}$ as a function of
  $\cos\theta^{\rm c.m.s.}_{D^+}$ where the thick line is the mean
  value of $A_{CP}$ while the hatched band is the $\pm1\sigma_{\rm
  total}$ interval, where $\sigma_{\rm total}$ is the total
  uncertainty.}
\label{FIG:ACPKSPI}
\end{figure}

Figure~\ref{FIG:ACPKSPI} shows invariant masses of $D^{\pm}\rightarrow
K^0_S\pi^{\pm}$ together with the fits that result in $\sim$1.74M
reconstructed decays and the measured $A_{CP}$ in bins of the polar
angle of $D^+$ momentum at the c.m.s. From the right plot in
Fig.~\ref{FIG:ACPKSPI}, we obtain $A^{D^+\rightarrow
  K^0_S\pi^+}_{CP}=(-0.363\pm0.094\pm0.067)\%$ which shows 3.2$\sigma$
deviations from zero. This is the first evidence for $CP$ violation in
charm decays from a single decay mode while the measured asymmetry is
consistent with the $A^{\bar{K}^0}_{CP}$. After subtracting experiment
dependent $A^{\bar{K}^0}_{CP}$~\cite{GROSSMAN_NIR}, the $CP$ violation
due to change in charm, $A^{\Delta C}_{CP}$, is measured to be
$(-0.024\pm0.094\pm0.067)\%$~\cite{KSPIPRL}.
\begin{figure}[htbp]
\mbox{
  \includegraphics[width=0.4\textwidth, height=0.35\textwidth]{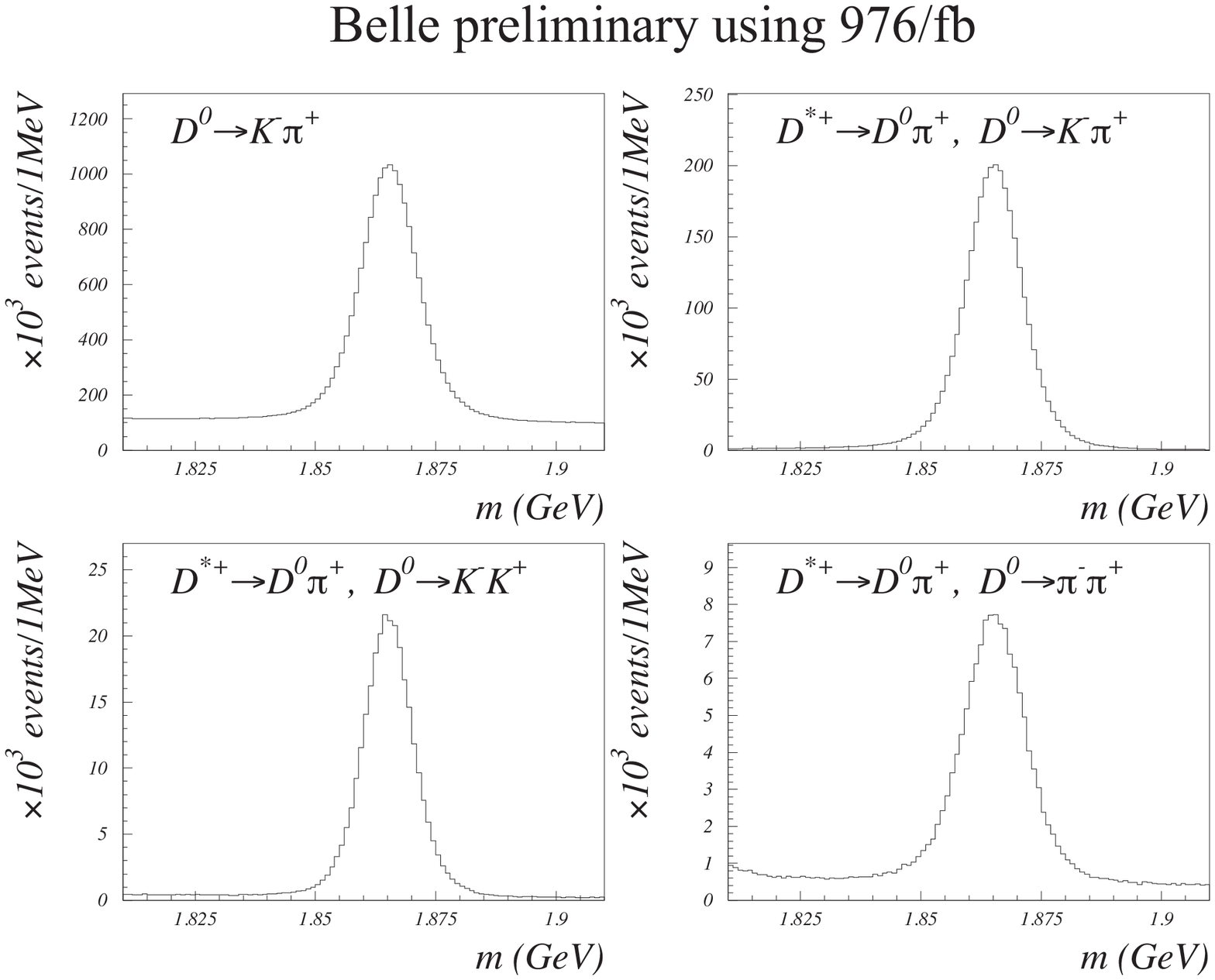}
}
\mbox{
  \includegraphics[width=0.6\textwidth, height=0.35\textwidth]{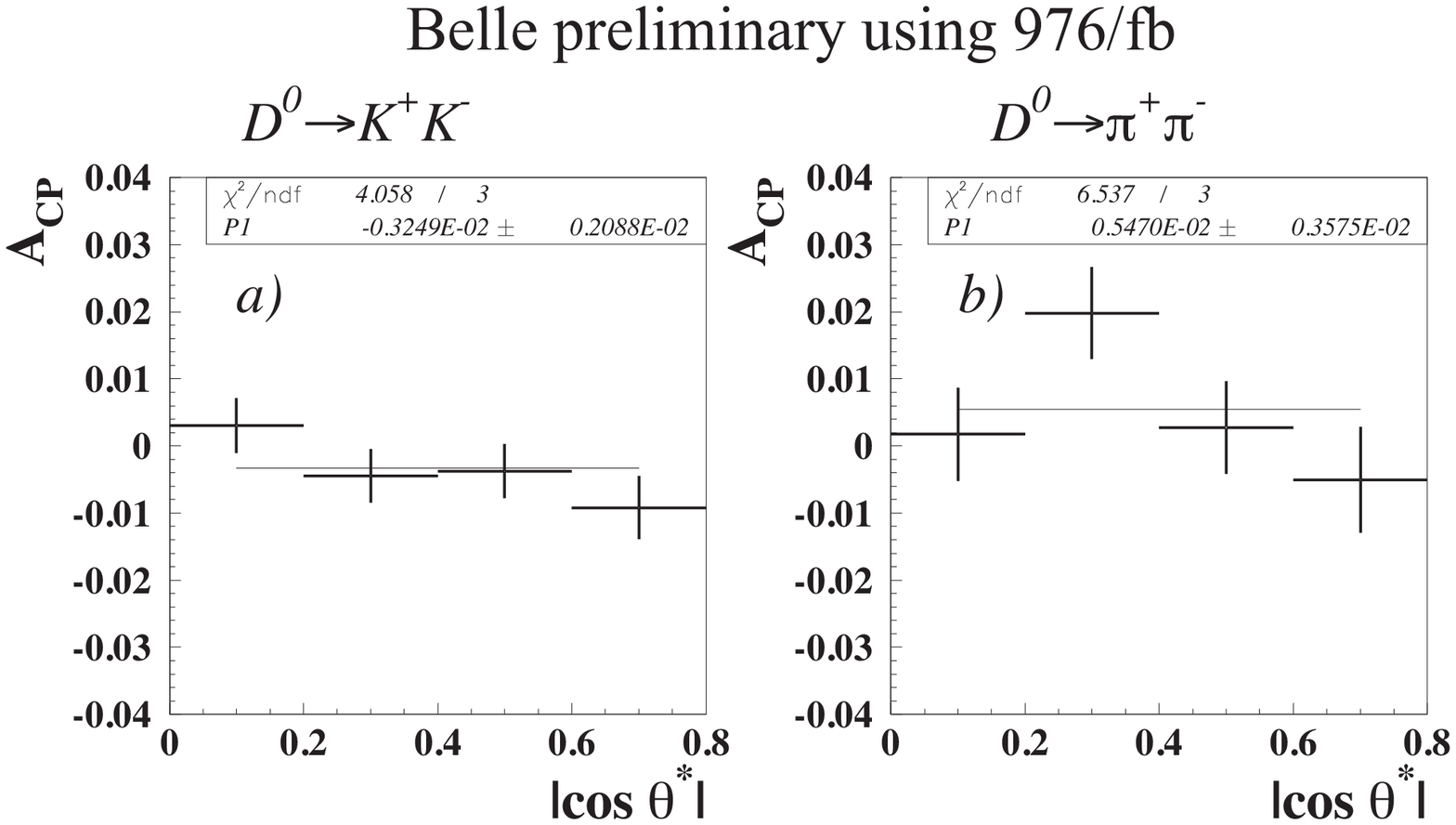}
}
\caption{Left four plots show reconstructed signal distributions
  described in the text and right two plots show preliminary results
  of $A_{CP}$ as a function of the polar angle of $D^{*+}$ momentum at
  the c.m.s.}
\label{FIG:ACPD0hh}
\end{figure}

Figure~\ref{FIG:ACPD0hh} shows reconstructed signal distributions
showing 14.7M $D^0\rightarrow K^-\pi^+$, 3.1M $D^{*+}$ tagged
$D^0\rightarrow K^-\pi^+$, 282k $D^{*+}$ tagged $D^0\rightarrow
K^+K^-$, and 123k $D^{*+}$ tagged $D^0\rightarrow\pi^+\pi^-$ on top of
the high signal purities, respectively, and the measured $A_{CP}$ in
bins of the polar angle of $D^{*+}$ momentum at the c.m.s. From the
right plot in Fig.~\ref{FIG:ACPD0hh}, we obtain
$A^{KK}_{CP}=(-0.32\pm0.21\pm0.09)\%$ and
$A^{\pi\pi}_{CP}=(+0.55\pm0.36\pm0.09)\%$ where the former shows the
best sensitivity to date. From the two measurements, we obtain $\Delta
A^{hh}_{CP}=(-0.87\pm0.41\pm0.06)\%$ which shows 2.1$\sigma$
deviations from zero and supports recent LHCb and CDF
measurements~\cite{ACP_LHCB, ACP_CDF}. By combining LHCb, CDF, and
Belle results, the average of $\Delta A^{hh}_{CP}$ becomes
$(-0.74\pm0.15)\%$.

\begin{figure}[htbp]
\begin{center}
  \includegraphics[width=0.7\textwidth, height=0.5\textwidth]{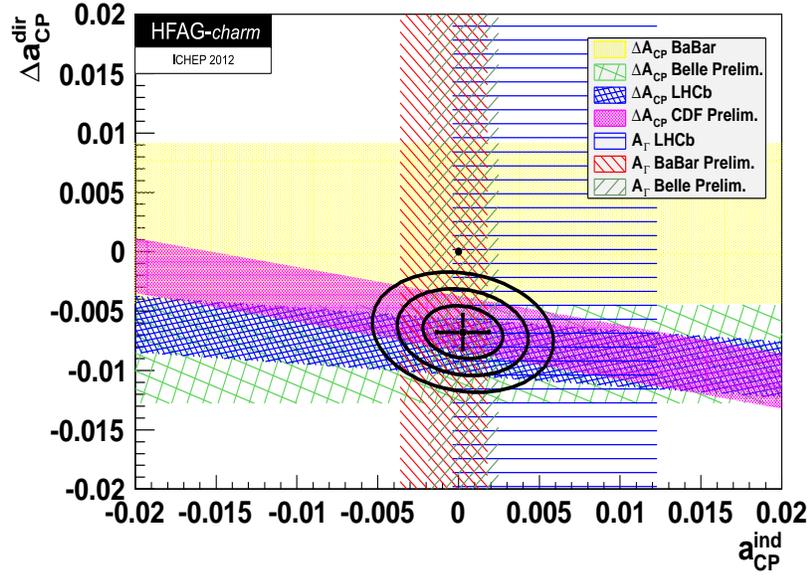}
\caption{$\Delta A_{CP}$ and $A_{\Gamma}$ fit from HFAG.}
\label{FIG:ACPhhICHEP}
\end{center}
\end{figure}

With a help from Marco Gersabeck from Heavy Flavor Averaging Group
(HFAG), Fig.~\ref{FIG:ACPhhICHEP} shows $\Delta A_{CP}$ and
$A_{\Gamma}$ fit reflecting the new Belle results reported in this
presentation and results in $\Delta a^{\rm
  dir}_{CP}=(-0.678\pm0.147)\%$ and $a^{\rm
  ind}_{CP}=(+0.027\pm0.163)\%$~\cite{HFAG}.

In summary, we observe evidence for $CP$ violation in the decay
$D^+\rightarrow K^0_S\pi^+$ where the evidence is consistent with the
expected $CP$ violation due to $K^0-\bar{K}^0$ mixing. No evidence for
$CP$ violation in $D^0\rightarrow h^+h^-$ is observed and the $\Delta
A^{hh}_{CP}$ is measured to be $(-0.87\pm0.41\pm0.06)\%$.


\begin{thebibliography}{99}
\bibitem{CC}
Throughout this Letter the charge conjugate decay mode is also implied
unless stated otherwise.
\end{thebibliography}
\end{document}